\documentclass[12pt,preprint]{aastex}

\def\chandra{{\it Chandra}}

\def\hst{{\it HST}}

\def\rosat{{\it ROSAT}}

\def\vla{{\it VLA}}

\def\vlbi{{\it VLBI}}
\def\glast{{\it GLAST}}

\def\flux{erg cm$^{-2}$ s$^{-1}$}
\def\nh{cm$^{-2}$}
\def\arcsec{$^{\prime\prime}$}
\def\deg{$^{\circ}$}

\def\ltsima{$\; \buildrel < \over \sim \;$}
\def\simlt{\lower.5ex\hbox{\ltsima}} 
\def\gtsima{$\; \buildrel > \over \sim \;$}
\def\simgt{\lower.5ex\hbox{\gtsima}} 

\def\s5{S5~2007+777}

\begin{document}

\title{A kpc-scale X-ray jet in the BL Lac source S5~2007+777}
\author{Rita M. Sambruna, Davide Donato, C.C.Cheung}
\affil{NASA/GSFC, Code 661, Greenbelt, MD 20771 (rms@milkyway.gsfc.nasa.gov)}

\author{F. Tavecchio and L. Maraschi}
\affil{Osservatorio Astronomico di Brera, via Brera 28, 20121 Milano, Italy}

\begin{abstract}

X-ray jets in AGN are commonly observed in FRII and FRI radiogalaxies,
but rarely in BL Lacs, most probably due to their orientation close to
the line of sight and the ensuing foreshortening effects. Only three
BL Lacs are known so far to contain a kpc-scale X-ray jet. In this
paper, we present the evidence for the existence of a fourth extended
X-ray jet in the classical radio-selected source \s5, which for its
hybrid FRI/II radio morphology has been classified as a HYMOR (HYbrid
MOrphology Radio source). Our \chandra\ ACIS-S observations of this
source revealed an X-ray counterpart to the 19\arcsec-long radio
jet. Interestingly, the X-ray properties of the kpc-scale jet in \s5\
are very similar to those observed in FRII jets. First, the X-ray
morphology closely mirrors the radio one, with the X-rays being
concentrated in the discrete radio knots. Second, the X-ray continuum
of the jet/brightest knot is described by a very hard power law, with
photon index $\Gamma_x \sim 1$, although the uncertainties are
large. Third, the optical upper limit from archival \hst\ data implies
a concave radio-to-X-ray SED. If the X-ray emission is attributed to
IC/CMB with equipartition, strong beaming ($\delta$=13) is required,
implying a very large scale (Mpc) jet. The beaming requirement can be
somewhat relaxed assuming a magnetic field lower than
equipartition. Alternatively, synchrotron emission from a second
population of very high-energy electrons is viable. Comparison to
other HYMOR jets detected with \chandra\ is discussed, as well as
general implications for the origin of the FRI/II division.

\end{abstract}


\section{Introduction}

Multiwavelength imaging spectroscopy of kpc-scale jets in AGN provides
direct clues to their structure and emission processes. Traditionally,
the information about jet structure first became available in the
radio band where the highest angular resolutions are achieved (Bridle
\& Perley 1984). These observations established that jets may extend
on quite substantial scales, with components being ejected from the
core and propagating at relativistic velocities; the radio emission is
understood as synchrotron radiation from relativistic electrons
accelerated via shocks or turbulence.

Optical observations from the ground and then with \hst\
established that many jets emit at these wavelengths, implying energies
$\gamma\sim 10^5$ and thus acceleration events far from the black
hole(Macchetto et al. 1994). The advent of the \chandra\ X-ray
Observatory and its discovery of ubiquitous X-ray emission from radio
jets has provided a further tool in the study of these systems.

Today, after almost a decade of \chandra\ observations, it is clear
that jet morphologies and spectral properties at X-rays are a function
of their power (see Sambruna, Maraschi, \& Tavecchio 2008 for a recent
review). Powerful jets hosted by FRII radio galaxies are generally
narrow and long, extending up to Mpc scales; their emission is
concentrated in discrete knots with good spatial correspondence
between the radio and the X-rays at least in the inner regions
(Sambruna et al. 2004, S04 in the following; Marshall et
al. 2005). The SEDs of individual bright knots often show two
components: a synchrotron one extending from radio to optical-UV, and
a second one uprising in the X-rays whose origin is still debated. A
popular scenario is inverse Compton on the CMB photons (IC/CMB), which
implies relativistic motion on large scales (S04).

Low-power \chandra\ jets are found in FRIs (Worrall, Birkinshaw, \&
Hardcastle 2001). They are shorter than their powerful counterparts,
terminating a few kpc from the core where the radio emission
``flares'' out in a fan-like structure. The X-ray flux lies on the
extrapolation of the synchrotron radio-to-optical emission and its
soft (photon index $\Gamma_X$ \gtsima 2) spectrum implies a cutoff in
or below the X-ray band. Remarkably, jets of intermediate powers
exhibit intermediate properties between FRI and FRIIs (Birkinshaw,
Worrall, \& Hardcastle 2002, Pesce et al. 2001, Sambruna et
al. 2007). However, very little is known on the large-scale X-ray
emission from jets in BL Lacs, the more aligned versions of FRI
sources. This is likely due to foreshortening effects associated to
the close alignment of the jet, which prevent the finding of
relatively extended and bright radio jets suitable for \chandra\
studies. One exception is \s5, which exhibits a bright 19\arcsec-long
radio jet with multiple knots and a hybrid FRI-II morphology (see
below). We observed \s5\ with \chandra\ in AO5 and detected the X-ray
counterpart of the radio jet, as discussed in this paper.

The paper is structured as follows. The source properties are
described in \S~2, and the observations in \S~3. The results of the
imaging and spectroscopy analysis for the core and the jet are given
in \S~4, while discussion and conclusions follow in \S~5. Throughout
this paper, a concordance cosmology with H$_0=71$ km s$^{-1}$
Mpc$^{-1}$, $\Omega_{\Lambda}$=0.73, and $\Omega_m$=0.27 (Spergel et
al. 2003) is adopted. With this choice, 1\arcsec\ corresponds to
4.8~kpc. The energy spectral index, $\alpha$, is defined such that
$F_{\nu} \propto \nu^{-\alpha}$.

\section{The source} 

The core of \s5\ is a classical radio-selected BL Lac from the 1 Jy
sample of Stickel et al. (1995), and a superluminal VLBI source
(Witzel et al. 1988; Gabuzda et al. 1994). Its redshift, $z=0.342$,
was derived from weak [OII] and [OIII] emission lines (Stickel et
al. 1989). Variability on time scales of a few days was observed from
the core at 100$\mu$m, and at radio and optical frequencies (Peng et
al. 2000).  Previous \rosat\ observations of the core showed a
power-law emission with $\Gamma_X=1.66$ (Urry et al. 1996), while
there is no detection at gamma-rays with EGRET (Hartman et al. 1999).  

\s5\ has a core-dominated radio morphology in \vla\ maps at
1.5 GHz (Antonucci et al. 1986), and on sub-pc scales with the
\vlbi\ (Perez-Torres et al. 1994). A prominent edge-brightened lobe is
present on the Eastern side of the nucleus at $\sim$ 11\arcsec,
whereas the Western side exhibits an 19\arcsec-long, knotty jet (see
Fig.~1) which ends with no clear hot spot. The radio morphology of
\s5\ - with an edge-brightened lobe on the Eastern side and a core-jet
morphology on the Western side - led to its classification as a HYMOR
(HYbrid MOrphology Radio source; Gopal-Krishna \& Wiita 2000), a
relatively rare class of sources. Their existence has been taken to
argue in favor of the role of the ambient gas in determining the
FRI-II division, as opposed to purely the properties of the central
engine.

Indeed, the core and jet 4.9~GHz luminosities in \s5\ are $\sim 5
\times 10^{33}$ and $\sim 10^{32}$ ergs/s/Hz, respectively. 
The magnitude of the host galaxy of \s5\ is M$_R=-23.89$ (Urry et
al. 2000). Thus, \s5\ lies at the boundary between FRIs and FRIIs in
the revised Owen \& Ledlow diagram (Owen \& Ledlow 1994) in Figure~7
of Urry et al. (2000).

We derived limits to the inclination of the radio jet using available
radio information.  \s5\ is a superluminal VLBI source (Witzel et
al. 1988; Gabuzda et al.  1994) with a single jet feature observed in
the 1980's moving at 4.7c (converting to our adopted cosmology). More
recent VLBA observations (Homan et al. 2001; Kellermann et al. 2004)
show at most, subluminal motions within the first 2 mas of the jet. As
discussed in Homan et al. (2001), the earlier observations probed
larger scale emissions and are not necessarily contradictory to the
newer observations.

The superluminal feature constrains the VLBI scale jet to be aligned
within $\sim$24\deg\ to our line of sight. The parsec-scale jet
(position angle, PA = -81\deg\ to -100\deg; Gabuzda et al. 1994) is
aligned to within 20\deg--30\deg\ (projected) of the X-ray detected
kpc-scale jet knots (PA$\sim$ -100\deg\ to -110\deg; Figure~1). The
intrinsic bend required to produce the projected misalignments is less
than (20--30\deg)x sin(24\deg) = 8\deg--12\deg. Therefore, the
kpc-scale jet is aligned to $<$ 32\deg--36\deg of our line of
sight. This limit on the inclination angle implies a relatively large
deprojected jet length, $>$150 kpc.

\section{Observations}

\subsection{Chandra} 

\chandra\ observed \s5\ on May 23, 2005. The live time was 36~ks 
yielding 33~ks of useful exposure time after removing flaring
background events that occured during the observations. The background
light curves were extracted from source-free regions on the same chip
of the source. The data were analyzed using standard screening
criteria. In particular we used the version 3.3 of the
\verb+CIAO+ software package and the calibration files provided
by the \chandra\ X-ray Center (\verb+CALDB v.3.2.1+).

The source was observed at the aim point of the S3 chip of the ACIS-S
detector. Since it was expected to have a bright core, the 1/2
subarray mode (frame time of 1.8 s) was used to reduce the effect of
the nuclear pileup. In a circular region of radius 1\arcsec\ the
counts rates of the core of \s5\ is 0.210 $\pm$ 0.003 c/s. Despite the
precautions, according to \verb+PIMMS+ the estimated pileup percentage
is non-negligible, 22\%, yielding an intrinsic (pileup-corrected)
net count rate of 0.34 c/s. The X-ray image shows a readout streak on
the S3 chip that has been removed by the \verb+destreak+ tool, for
imaging purpose. The core flux and spectrum was extracted using a
circular region centered on the pixel with the highest counts with
radius 3\arcsec. The background was chosen in a \verb+panda+ region on
the same CCD, centered on the source and free of the jet and
serendipitae sources.

For the detected knots, fluxes and spectra were extracted from
ellipses centered on the radio (1.49~GHz) positions and with axis
lengths as listed in Table~1. The corresponding aperture corrections
are in the range 1.01--1.1. In all cases the background was evaluated
in \verb+panda+ regions centered on the nucleus and with inner and
outer radii such as to match the width of the source extraction
region. The choice of this background region ensures that the
contribution of the wings of the PSF as a function of azimuth are
averaged. Figure~1 shows the ACIS image of the source.  The net count
rates of the knots are reported in Table~1. The uncertainties on the
X-ray counts, $\sigma_N$, were calculated according to the formula
$\sigma_N=[(\sigma_S)^2+(\sigma_B)^2]^{1/2}$, where $\sigma_S$ and
$\sigma_B$ are the uncertainties of the source and the re-scaled by
area background, respectively. In the regime of low counts, to
evaluate the uncertainties we adopted the formula in Gehrels (1986):
$\sigma = 1+(S+0.75)^{1/2}$.  

The ACIS spectra of the core and jet have been analyzed within
\verb+XSPEC v.11.3.2+.  For the jet, spectra were extracted for knot
K8.5, where 40 counts were obtained, and for the full jet (101
counts). For the latter, a rectangular box of dimensions 15.5\arcsec
x3.5\arcsec\ was used. The spectrum of the core and full jet were
grouped so that each new energy bin had at least 20 counts to enable
the use of the $\chi^{2}$ statistics.  Errors quoted throughout are
90\% for one parameter of interest ($\Delta\chi^2$=2.7). The spectrum
of knot K8.5, where less than 100 counts were detected, was not
rebinned and it was analyzed with the C-statistic.

Comparing the jet's total counts to the core's shows that the
jet accounts for only 1\% of the total X-ray emission from \s5. Thus,
previous X-ray observations at lower angular resolution were dominated
by the unresolved nuclear source emission. 

\subsection{HST} 

To constrain the level of optical flux from the jet, we observed \s5\
with \hst\ on May 29, 2005 (GO program 10343). We selected the
Advanced Camera for Surveys (ACS; Sirianni et al. 2005) because of the
observing efficiency of the WFC camera, and utilized the F814W filter
(broad I; pivot wavelength of 8060\AA). Over the one orbit allocation,
a total exposure of 2763 sec was obtained, split evenly between 3
individual exposures. We additionally obtained from the
\hst\ archive a WFPC2 F702W (440 sec; pivot wavelength 6917\AA)
snapshot image of \s5\ (Scarpa et al. 2000; Urry et al. 2000).

We found no obvious optical counterparts to the radio/X-ray jet knots
in the \hst\ images pre- and post-subtraction of the central
quasar. Although there are apparent optical excesses in the vicinity
of the inner knots (Figure~1), the count rate measured in 15$\times$15
pixel boxes centered on the radio peaks are within 1$\sigma$ of the
average count rates in adjacent apertures. To derive optical upper
limits, we converted these measurements of the fluctuations in the
average count rates to flux densities using the inverse sensitivity
measurements contained in the PHOTFLAM keyword.

\subsection{Radio}

To study the radio jet, we obtained matched resolution ($\sim$1$"$)
VLA datasets at 1.49 GHz (A-configuration) and 4.86 GHz
(B-configuration) from the NRAO archive.  The 1.49 GHz data is from
Feb 1985 (program AL41; previously published in Antonucci et al. 1986)
and has a total exposure of $\sim$35 min.  The ($u,v$) coverage was
predominantly in the East-West directions, amounting to an elongated
North-South beam and resulting in relatively bright artifacts in this
direction like the feature just South of the radio core.

The 4.86 GHz data were obtained as part of a variability monitoring
campaign of \s5\ in 1997 (Peng et al. 2000). To produce the image, we
combined their observations from March 1--7, amounting to about an
hour of total exposure time. Peng et al. (2000) found two epochs where
the flux densities differed by more than 1$\sigma (=2$\%) of the
average during this period (but only up to 2--3$\sigma$).  We found
little difference in the images produced excluding these two exposures
and that produced from all of the data so retained the full dataset.

\section{Results} 

The 0.3--8~keV image of \s5\ is shown in Figure~1, together with the
radio and optical data. A prominent jet is present in the X-ray image, 
in addition to the unresolved core. Below we discuss separately the X-ray
data for the core and the X-ray and multiwavelength observations of the jet.

\subsection{The Core} 

Inspection of the ACIS-S images suggests that the core is unresolved
and consistent with a point source. For a more quantitative analysis
we extracted the radial profiles of the core region, using a series of
concentric annuli and \verb+panda+ regions centered on the core and
extending out to 360\arcsec\ on the ACIS CCD, with the jet
excised. For the extraction procedure, see Donato et al. (2004). The
jet and field point sources contributions were excluded from the
extraction regions. The ACIS spectrum of the core (see below) was
assumed.

The radial profiles are well fitted by the instrumental PSF, with no
need for extra components. In particular, there is no evidence for
excess emission over the wings of the PSF which could be attributed to
diffuse X-ray emission, contrary to other sources at similar redshifts
(Sambruna et al. 2007). Thus, any diffuse circumnuclear emission is
confined within 10\arcsec, or 48~kpc. 

No flux variability is detected from the core within the 33~ks
exposure. A total of 7,800 counts were measured in 0.3--8~keV,
sufficient to perform spectral analysis. The ACIS-S spectrum of the
core is well fitted by a single power law with Galactic
N$_H$=8.58$\times10^{20}$ \nh. To account for residual pile-up (see
\S~3.1) the multiplicative component \verb+pileup+ was included in the
model. For the latter, the following parameters (fixed) were assumed:
\verb+fr_time+=1.8, \verb+max_ph+=5, \verb+g0+=1, and
\verb+nregions+=1. The free parameters were
\verb+psffrac+=0.96$^{+0.03}_{-0.30}$, $\alpha=0.49^{+0.51}_{-0.16}$. 

For the power law, the fitted photon index is $\Gamma=1.98 \pm 0.23$
and the observed flux is F$_{0.3-8~keV}\sim 2.5 \times 10^{-12}$
\flux. The slope is consistent within the large uncertainties with the
\rosat\ value (Urry et al. 2006), while the extrapolated 0.3--8~keV
flux from the latter data is within a factor 2.

\subsection{The Jet}

\subsubsection{The Images} 

Figure~1 shows a montage of the images of the jet at different
wavelengths, in order of decreasing wavelength. The 1.49~GHz radio
contours from the first image at 1.5\arcsec x0.9\arcsec\ resolution
are overlaid on all the following ones; the 4.86 GHz image was
convolved with the same beam. The \chandra\ image in the energy band
0.3--8~keV was rebinned by a factor 0.5 and adaptively smoothed with
\verb+csmooth+ in \verb+CIAO+ using a Gaussian function of standard 
deviation 2 pixels ($\sim$ 0.5\arcsec). Assuming a typical value for
the ACIS standard deviation of 0.8\arcsec, the final resolution is
$\sim$ 1.2\arcsec, comparable to the radio. To obtain a similar
resolution in the optical, the ACS data were smoothed with
\verb+fgauss+ in \verb+FTOOLS+ with a standard deviation of 7 pixels, 
or $\sim$ 0.35\arcsec, with a corresponding FWHM $\sim 0.9$\arcsec.

The most striking result from Figure~1 is the detection of an X-ray
counterpart to the radio jet, with very similar morphology at the two
wavelengths. Such a match between radio and X-ray morphology is rarely
found in FRIs but can be observed in powerful FRIIs (e.g., 1354+195;
S04; Schwartz et al. 2007). However, in a significant number of FRII
jets the X-ray-to-radio flux ratios decrease along the jet (S04;
Sambruna et al. 2006). There is no detection of the jet in the
optical.

The radio emission of the jet consists of several discrete knots
extending SW of the core. In this paper we adopt the conventional
nomenclature for the knots Kx.x where x.x is the distance in arcsec of
the knot from the core. A total of 5 radio knots are present, for an
overall extent of 19\arcsec, with the brightest radio knot being
located mid-stream at $\sim$ 8.5\arcsec. At 1.49~GHz, the jet extends
rather linearly until knot K11.1, after which a bend is present in the
W direction with a position angle change of $\Delta PA\simeq 20$\deg. 

In the ACIS-S image the radio jet exhibits X-ray emission from every
knot, fading at the end of the radio jet. The brightest X-ray
counterpart, with 40 net counts or 43\% of the total jet counts
(Table~1), is knot K8.5, which is also the brightest in the radio. The
X-ray counts from the knots  are listed in Table~1. 

A weak radio counterlobe is present in the radio images of \s5\ (not
shown here). There is no X-ray or optical detection of neither the lobe
or its hotspot.

To better quantify the jet properties we extracted profiles along the
main jet axis (longitudinal profiles) at three wavelengths: radio
at both 1.49~GHz and 4.86~GHz, and X-rays (0.3--8~keV). The images
were reprocessed to ensure similar resolutions (0.9\arcsec) at all
wavelengths as explained above. The profiles were extracted by
collapsing the flux of the jet onto a box 19\arcsec-long and
1.5\arcsec-wide. The profiles of the jet are shown in Figure~2. 

The radio core is bright and the \vla\ images are dynamic range
limited. With the overall faintness of the jet, the details of the
fainter radio features are uncertain as evident from the slightly
differing profiles between the two radio images. In this respect, the
knot positions are based on the \chandra\ data, and regions where
there are apparently multiple radio knots, like K15.9, are defined as
a single region. The low signal-to-noise ratio \chandra\ data for the
jet makes it difficult to make definitive statements about possible
radio/X-ray offsets for all but the brightest knot, K8.5. Here, there
is a possible centroid shift by 0.5\arcsec\ with the X-ray peak
downstream of the radio one.

\subsubsection{The spectra} 

ACIS spectra were extracted for the full jet and for the brightest
knot K8.5, following the procedure described in \S~2.1. The ACIS
spectra of both the jet and K8.5 are consistent with a single power
law with Galactic column density; the parameters are listed in
Table~2. The fitted photon indices are hard, $\Gamma_X \sim$ 1.0,
albeit within large uncertainties. The jet emission is thus harder
than the nuclear emission, similar to what is observed in the
intermediate FRI/II 3C~371 and PKS~2201+044 (Sambruna et al. 2007).

We also tried a thermal model fit to the ACIS spectra of the jet. For
the full jet, a thermal model is equally acceptable, but at the fitted
temperature, kT=64 keV, it effectively mimicks a power law.  Similar
results are found for knot K8.5. 

The observed 0.3--8~keV fluxes for the core and the jet are listed in
Table~2. Table~1 reports the monochromatic flux densities at 1 keV for
the individual knots.  The flux densities were calculated with
\verb+PIMMS+ using the knot's count rate in column 2 and assuming a
power-law spectrum with photon index $\Gamma=1.02$ and Galactic
N$_H$. All fluxes are background-subtracted and corrected for finite
aperture effects.

Also listed in Table~1 are the radio energy indices $\alpha_R$ for
each knot calculated between 1.49 and 4.86~GHz. With $\sim$10-15$\%$
errors in the individual flux densities, the errors in the two point
spectral indices are $\sim$0.15-0.2 (larger for the fainter
features). This means the spectrum of the radio jet is basically
constant within the joint errors.

Using the flux densities at 4.86~GHz and 1~keV we calculated the
radio-to-X-ray index, $\alpha_{rx}$. The latter is reported in the
last column of Table~1. For all knots $\alpha_{rx} \sim 0.8$, fully
consistent with the distribution observed for FRII jets (S04). There
are no variations of $\alpha_{rx}$ along the jet of \s5, at odds with
most FRIIs (e.g., Figure~5 in S04); whereas in \s5\ the ratio is
remarkably constant ($\alpha_{rx}$ within 0.1 of 0.8).

\section{Discussion}  
  
\subsection{X-ray emission process}   
  
With its one-to-one correspondence between the X-ray and radio
emission, the \chandra\ jet of \s5\ strongly resembles that of
powerful FRIIs (S04). As in FRIIs, this morphological similarity
provides first clues to the origin of the high-energy emission. A
natural possibility is that the X-rays are produced via inverse
Compton scattering of seed photons off the same electrons responsible
for the synchrotron radio emission.  If the X-rays were the
high-energy tail of the radio synchrotron emission, the knot intensity
would be expected to decrease, and the continuum to soften, with
respect to the radio emission, unless reacceleration occurs throughout
the volume. Indeed in 3C~371 and PKS~2201+041 where the synchrotron
model provides an adequate description of the SEDs the X-ray spectra
clearly steepen along the jet (Sambruna et al. 2007).

To better quantify the properties of the X-ray jet of \s5\ and derive
physically interesting parameters, we assembled its Spectral Energy
Distribution (SED) from radio to X-rays using the results in Table~1
and 2. The SEDs of the various knots are shown in the bottom panel of
Figure~3. For knot K8.5 we also plot the X-ray spectrum and a
3$\sigma$ upper limit to the optical emission derived from the ACS
data. Taking into account the upper limit in the optical, the SED for
knot K8.5 appears concave; furthermore, the X-ray continuum appears
hard, implying that a large fraction of the emission occurs above the
X-ray energy band and is currently unseen. 
  
The top panel of Figure~3 shows the SED of the core of \s5. The latter
was assembled using literature data (from NED) as well as the ACIS
spectrum from this work. The previous \rosat\ continuum is also
plotted for comparison (Urry et al. 1996), as well as the EGRET upper
limit to the GeV flux. The core optical-to-X-ray SED appears concave,
as commonly observed in intermediate-luminosity, low-energy peaked BL
Lacs. The X-ray emission from the core is weaker compared to the radio
by a factor 10. In the core SED the fraction of luminosity contained
in the low- and high-energy humps is estimated to be roughly similar,
while the jet is clearly dominated energetically by the high-energy
component.

To model the SEDs of both the core and jet, we assumed a synchrotron +
inverse Compton (IC) model. The basics of this model are described in
Maraschi \& Tavecchio (2003) and Tavecchio et al. (2000). The solid
lines in the panels of Figure~3 represent the closest attainable
fit. The corresponding parameters are reported in Table~3.  A
different origin of the seed photons for the IC process was assumed in
the core and the jet. 

In the core, seed photons are generally provided - in BL Lacs like
\s5\ - both by the synchrotron photons themselves (SSC) and ambient
photons external to the jet (EC). The lack of strong thermal features
in the optical, together with the EGRET upper limit and the slope of
the X-ray continuum, suggests a limited importance of the EC component
and thus we consider only SSC emission, with the result that the
high-energy component in the core SED cuts-off above a few GeV.  We
also choose to reproduce a relatively low state consistent with the IR
upper limits from IRAS, while the simultaneous observations of Peng et
al. (2000) refer to a higher state.  The similarity of the
multiwavelength jet to FRIIs suggests that, as in the latter sources,
the X-ray emission from the \s5\ jet could originate from IC
scattering of the CMB photons off the jet electrons (for criticisms to
this intepretation see Harris \& Krawczynski 2006). The hard X-ray
emission can be reproduced assuming that it belongs to the low-energy
tail of the IC/CMB component: this choice implies a relatively large
value for the minimum energy of the relativistic electrons,
$\gamma_{min}=70$, compared to the ``typical'' value
$\gamma_{min}=10-20$ for FSRQs (S04).

A Doppler factor of $\delta=13$ is derived with the IC/CMB model,
assuming equipartition between electrons and magnetic field. This
implies that the jet is observed under a rather small viewing angle,
$\theta \approx 4$\deg--5\deg, which, in turn, translates into a
deprojected length of the jet of $\approx 1$ Mpc. This length is
comparable to those of the FRII jets hosted by quasars (S04),
estimated through IC/CMB modeling. Considering that the jet of \s5\
has a clear FRII morphology and that the radio power of the jet
($5\times 10^{33}$ erg s$^{-1}$ Hz$^{-1}$) is not too far from that of
the FRII sources of S04 (clustering around $10^{34}$ erg s$^{-1}$
Hz$^{-1}$), a length of 1 Mpc could be in principle acceptable. A
possible way to reduce the value of the Doppler factor, thus
increasing the required viewing angle, is to relax the equipartition
condition, allowing for the electrons to dominate over the magnetic
field. For example, to get an inclination angle of 20\deg, the
electron density would be a factor $10^5$ larger than the magnetic
density. The IC/CMB model for the jet is plotted in Figure~3 as a
solid line.

An alternative possibility is that the X-ray emission is instead
produced by synchrotron radiation from a second population of
relativistic electrons with very high maximum energy. The latter could
coexist with the radio emitting electrons (e.g., Uchiyama et al. 2007)
or belong to another region of the jet (see e.g. Jester et al. 2006
for the jet of 3C~273). Stawarz \& Ostrowski (2002) proposed that a
turbulent acceleration mechanism operating in a boundary layer around
the jet could lead to a piled-up energy distribution, whose peaked
synchrotron emission can be quite hard (up to $\nu^{1/3}$). The
emission is expected to peak at X-rays if the characteristic velocity
of the magnetic turbulence is of the order of the Alfven velocity. A
synchrotron model along these lines is shown in Figure~3 (dashed line)
and the parameters are reported in Table~3. We assume two cospatial
electron populations. The first one (S1) follows a power law energy
distribution up to $\gamma _{\rm max}=4\times 10^5$ and is responsible
for the radio emission. The X-ray emission is reproduced assuming a
power-law electron distribution (S2) extending from $\gamma = 10^8$ to
$\gamma = 10^9$. The particle energy density of the latter is
negligible with respect to that of S1, which is in equipartition with
the the magnetic energy density.

Thus, the X-ray emission observed from the jet of \s5\ and others
(S04) could be explained by the synchrotron emission of high-energy
electrons piled-up in the jet boundary layer.  On the other hand, this
solution is largely underconstrained. Moreover, the good coincidence
between the radio and the X-ray emission suggests a close relation
between the radio emitting and X-ray emitting relativistic electrons
in this source, lending support to the IC/CMB interpretation.

An interesting feature of the IC/CMB model is that it predicts
relatively large emission at gamma-rays, with a very hard spectrum at
GeV energies, while the synchrotron model has a cutoff at a few MeV
(as discussed for 3C~273 by Georganopoulos et al. 2004). The exact
level and position of the IC/CMB peak depends on the value of maximum
energy of the relativistic electrons, which is not well constrained
due to the lack of data between the radio and optical band.  With the
conservative assumption adopted here that the peak of the synchrotron
component lies around $10^{12}$ Hz, the jet emission above a few GeV
is comparable to the predicted, softer emission from the
core. Conceivably future \glast\ observations of
\s5\ during core low-states could in principle discriminate the origin
of the jet X-ray flux on the basis of a detection at GeV energies with
a hard spectrum. However, at the predicted flux level, long
integration times are necessary for a detection, thus this test does
not look feasible for \s5\ until after a few years of the LAT
operations. A more promising source is 3C~273 (Georganopoulos et
al. 2004).

\subsection{Comparison with other HYMOR jets detected with \chandra}   
  
The radio morphology of \s5\ - with a lobe and hotspot on the Eastern
side and a core-jet morphology on the Western side - has led to its
classification as a HYMOR (HYbrid MOrphology Radio source;
Gopal-Krishna \& Wiita 2000). The hybrid morphology has been taken to
demonstrate that the origin of the FRI / FRII division is also to
ascribe to the environment, with gas distribution asymmetries on the two
sides of the nucleus being responsible for different jet fates
(Gopal-Krishna \& Wiita 2000). This would rule out models attributing
the origin of the dichotomy solely to different central engines or jet
composition in the two classes. 
   
Interestingly, from the list of 6 sources compiled by Gopal-Krishna \&
Wiita (2000) two other HYMOR jets were observed and detected with
\chandra. These are the z=0.055 BL Lac PKS~0521-365 
(Birkinshaw et al. 2002) and the z=0.240 BAL quasar PG~1004+130
(Miller et al. 2006).  The radio morphology of both sources is
strongly reminiscent of \s5, with a radio lobe + hotspot on one side
of the core and a jet without lobe/hotspot on the other. However, the
X-ray properties are somewhat different.

In PKS~0521--365 the X-ray observations reveal diffuse thermal
emission around the core on the host galaxy halo's scales (few
kpc). The X-ray jet emission peaks up-stream of the radio knot at a
$\sim$ few kpc from the core, and is described by a steep $\Gamma>$ 2
power law model indicative of synchrotron emission (Birkinshaw et
al. 2002). Overall, the X-ray jet and environment of PKS~0521--365
resemble strongly an FRI. Note that at the lower redshift of this
source we are probing relatively smaller scales than in \s5. Also,
since both weak broad and narrow optical emission lines are detected
in PKS~0521--365 over the variable non-thermal continuum (Danziger et
al. 1979), beaming is likely to be modest in this source. Indeed, a
fit of the core SED provides an inclination angle $\sim$ 30\deg\ (Pian
et al. 1996).
  
A more appropriate and interesting comparison is between \s5\ and
PG~1004+130 as their redshifts are similar. In the BAL quasar X-ray
emission from the SE radio jet was detected with \chandra\ (Miller et
al. 2006) at 8\arcsec\ (30 kpc) from the core. As noted by Miller et
al. (2006), the X-ray emission is aligned with the FRI jet but
displaced upstream with respect to the radio emission by about
5\arcsec\ which could indicate deceleration if the X-ray emission is
attributed to the IC/CMB process. On the contrary, in \s5\ the offset
between the X-ray and radio emission is negligible and if any, in the
opposite direction - the radio precedes the shorter wavelengths.
  
A possible difference between the two sources is the slope of the
X-ray continuum emission of the two jets. In PG~1004+130 the latter is
$\Gamma_{jet}=1.7$, while for \s5\ we found a flatter value (but
within large 1$\sigma$ uncertainties), $\Gamma_{jet}\simeq 1.0$. In
both jets, the optical upper limit, lying below the extrapolation from
radio to X-rays, implies a concave SED. Using the ratio of the core
radio to optical flux, Miller et al. derive an inclination angle
\gtsima 45\deg\ for PG~1004+130, and infer that beaming is not
affecting the jet properties. An IC/CMB fit to the SED yields
$\delta=3$ and an inclination angle $<$ 19\deg\ (Miller et
al. 2006). Based on this value for the angle, Miller et al. dismiss an
IC-CMB origin for the X-ray emission, and prefer to model the X-ray
emission with synchrotron from a second particle population in the
jet.

Note, however, that in the 4.9~GHz map shown by Gopal-Krishna \& Wiita
(their Fig.~1c) the radio jet of PG~1004+130 appears much more
prominent and better collimated than in a typical FRI suggesting that
some amount of beaming could be present. Indeed, the method used by
Miller et al. (2006) for estimating the inclination angle, based on
the core radio-to-optical luminosity ratio, R$_V$ (Wills \& Brotherton
1995) and core-to-lobe flux ratio, suffers from very large uncertainties.
We thus suggest that the IC/CMB model discussed by Miller et
al. (2006) implying modest beaming ($\delta \sim 3$) and a relatively
large angle, $\sim$ 20\deg, could be viable. Weak beaming is also
inferred from the X-ray properties of the jet in PKS~0521--365 (see
above). Thus, \s5\ stands out among the HYMOR jets probed by \chandra\
as the source with the largest beaming if the IC/CMB model is adopted.
This result is consistent with its core classification as a BL Lac object.

In this context, there are some interesting implications for the cause
of the FRI/II division. The general tendency of the X-ray emission
from bona fide FRII jets to fade before the end of the radio emission
can be naturally interpreted as deceleration of the jet bulk
motion. However, in the context of an IC/CMB origin of the X-ray
emission, if deceleration is due to entrainment of external
matter (Tavecchio et al. 2006; but see Hardcastle 2006), the length a
jet can travel before becoming subrelativistic depends on the
transported bulk kinetic energy. It follows that in similar gas
environments more powerful jets will remain relativistic over larger
scales.  Thus, a combination of interactions with ambient gas and jet
power are required to account for the different FRI/FRII morphologies,
confirming earlier findings (Owen \& Ledlow 1994). From this
perspective, it may not be coincidental that hybrid morphologies are
found at intermediate powers, when relatively small changes in the
environment can cause deceleration to set in at different scales.

\section{Conclusions} 
 
We have analyzed and discussed the kpc-scale radio and X-ray jet in
the classical radio-selected BL Lac \s5. Its properties appear
peculiar, both for the close correspondence of X-ray and radio
morphologies and for its hard X-ray continuum. Moreover, if an IC/CMB
origin for the jet X-ray emission is accepted, the jet intrinsic size
appears rather extreme. Relaxing the equipartition condition, however,
ameliorates this difficulty.

Among the HYMOR jets detected with \chandra, \s5\ stands out as
unusual in its similarity to powerful FRIIs, and in possibly requiring
the highest beaming. On the other hand, a synchrotron origin for the
X-ray jet, can not be ruled out on the basis of the present data.

\acknowledgements

The \vla\ is a facility of the National Radio Astronomy Observatory is
operated by Associated Universities, Inc. under a cooperative
agreement with the National Science Foundation. Based in part on
observations made with the NASA/ESA Hubble Space Telescope, obtained
from the data archive at the STScI.  STScI is operated by the
Association of Universities for Research in Astronomy, Inc. under NASA
contract NAS 5-26555. This research has made use of the NASA/IPAC
Extragalactic Database (NED) which is operated by the Jet Propulsion
Laboratory, California Institute of Technology, under contract with
the National Aeronautics and Space Administration. C.C.C. was
supported in part by an appointment to the NASA Postdoctoral Program
at the Goddard Space Flight Center, administered by Oak Ridge
Associated Universities through a contract with NASA.



\clearpage


\begin{figure}[h]
\centerline{\includegraphics[height=5.5in]{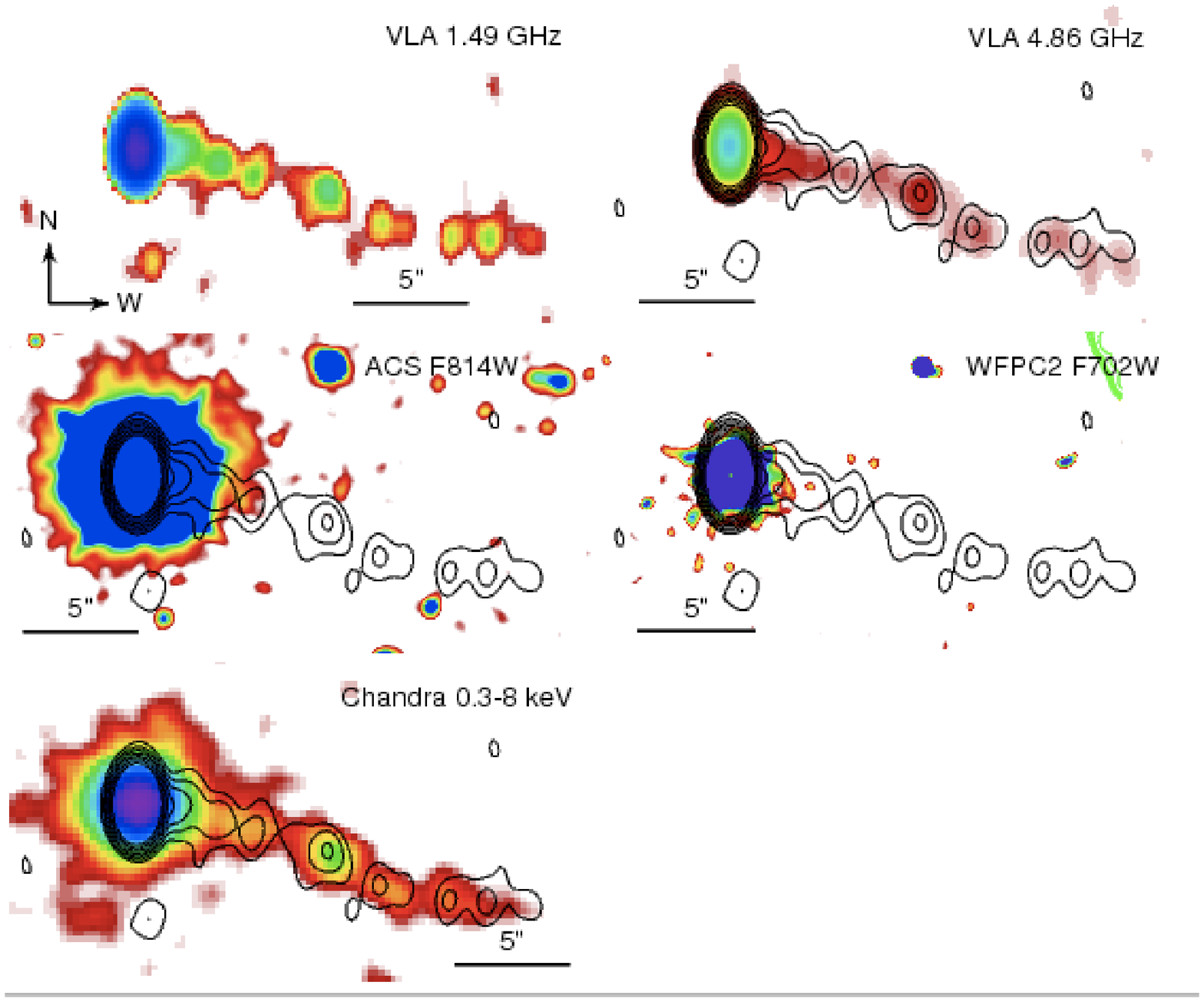}}
\caption{
Multiwavelength images of the jet of S5~2007+777. 
First row: \vla\ 1.49~GHz and 4.86~GHz;
Second row: \hst\ ACS (F814W) and WFPC2 (F702W); 
Third row: \chandra\ 0.3--8 keV.
In all cases, the 1.49~GHz radio contours are
overlaid on the image.
}
\end{figure}

\clearpage


\begin{figure}[h]
\centerline{\includegraphics[height=7.in]{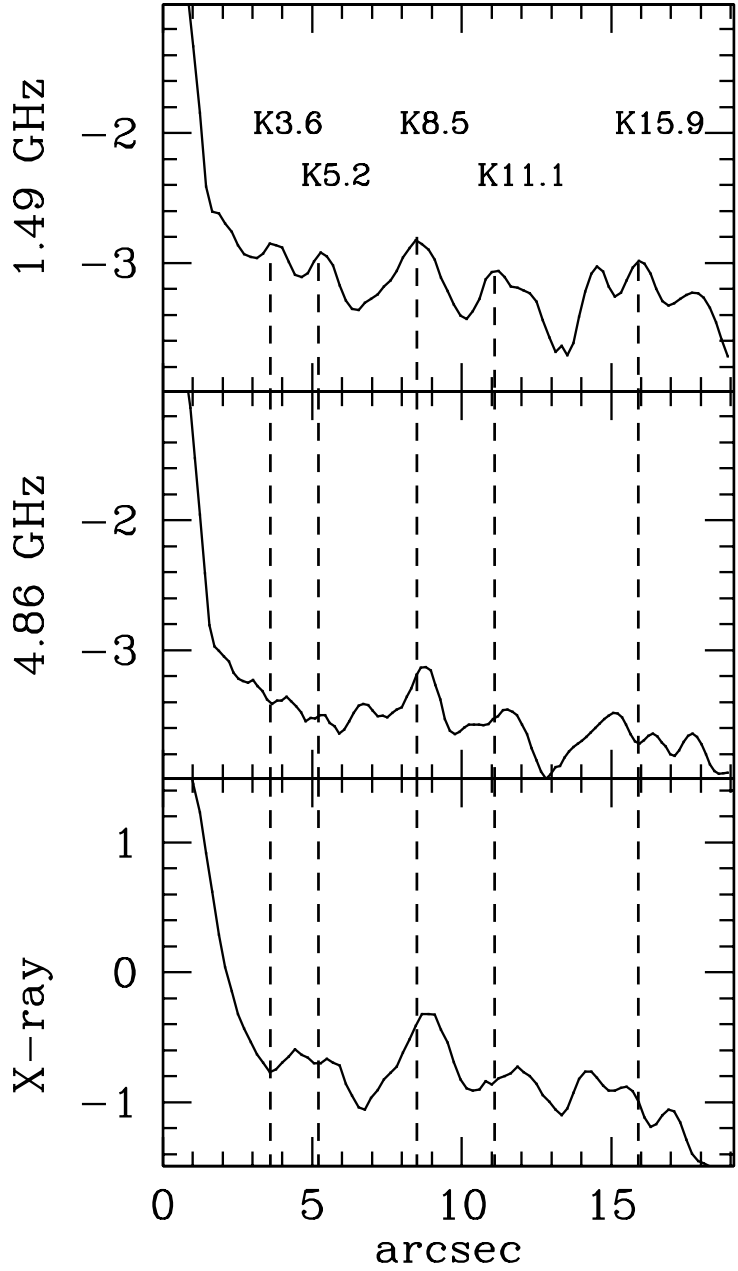}}
\caption{
Radial profiles of the jet. The y axis is the logarithm of the count
rates. The background is below the scale. The position of the radio
knots is labeled by the vertical dashed lines. Uncertainties are
\gtsima 30\% for the X-rays and 10--15\% for the radio.}

\end{figure}

\clearpage


\begin{figure}[h]
\centerline{\includegraphics[height=6.in]{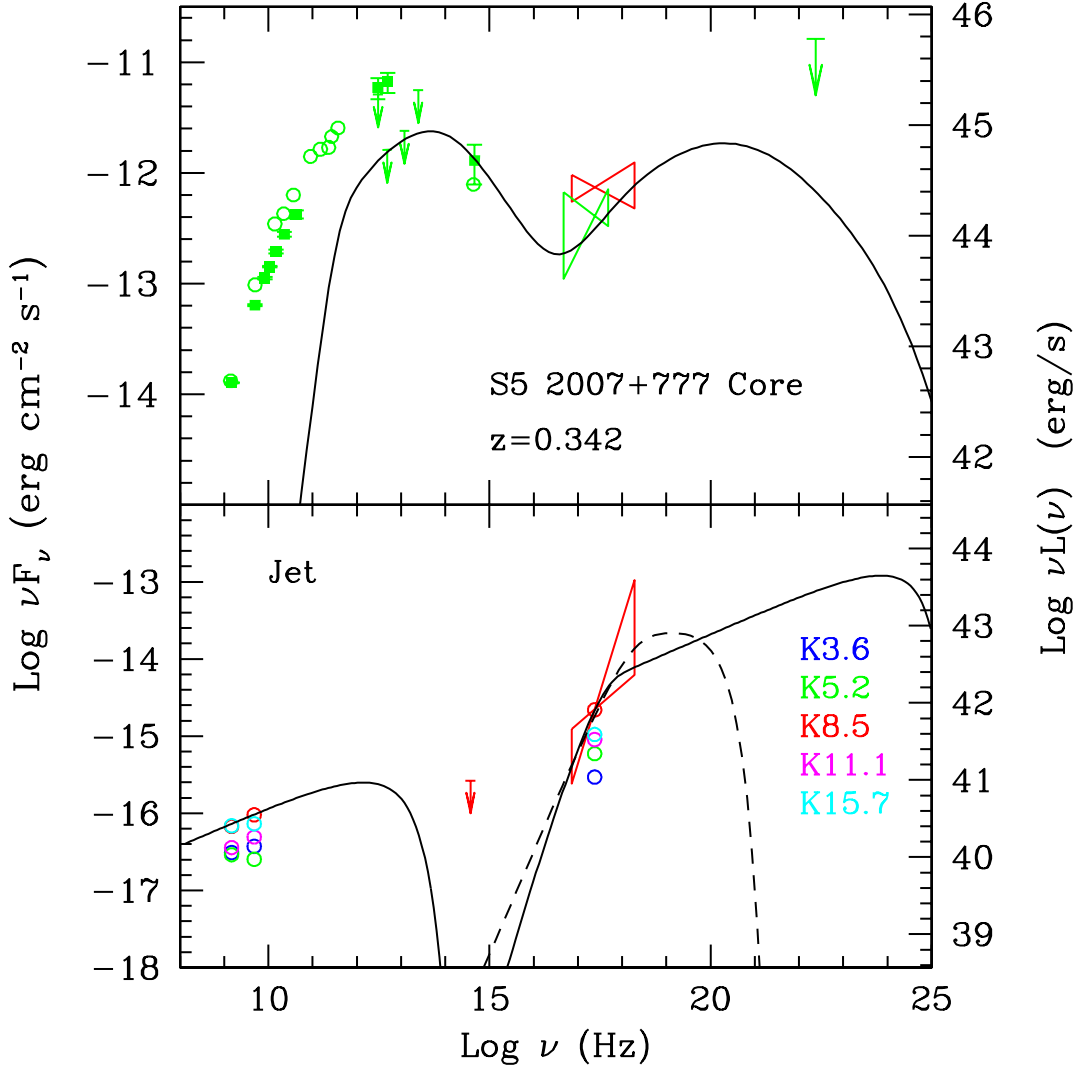}}
\caption{\footnotesize
Spectral Energy Distributions of the core (top) and of the jet
(bottom) for S5~2007+777. {\it Top panel:} The data for the core were
taken from NED (open symbols) while the filled squares are the
simultaneous measurements from Peng et al. (2000). The bowties at
X-rays represent the \rosat\ (Urry et al. 1996) and the \chandra\
(this work) observations. The solid line is a fit to the SED with the
SSC model (Table~3). {\it Bottom panel:} The SEDs for all the knots in
the jet are shown, using the data in Table~1 and 2. For the brightest
knot K8.5 the X-ray continuum is represented with a bowtie. Two models
are reported: the IC/CMB (solid line) and two synchrotron components
(dashed line). In the latter, the low-energy synchrotron hump overlaps
completely to the synchrotron part of the IC/CMB model. The parameters
for all models are reported in Table~3. }

\end{figure}

\end{document}